\documentclass[aps,pra,preprint,showpacs]{revtex4}
\usepackage{graphicx}
\usepackage{amssymb}

\begin{document}

\title{Modeling long-range memory trading activity by stochastic differential equations}

\author{V. Gontis }

\email{gontis@itpa.lt}

\homepage{http://www.itpa.lt/~gontis}

\affiliation{Institute of Theoretical Physics and Astronomy of Vilnius
University\\ A. Go\v{s}tauto 12, LT-01108 Vilnius, Lithuania }

\author{B. Kaulakys}

\affiliation{Institute of Theoretical Physics and Astronomy of Vilnius
University\\ A. Go\v{s}tauto 12, LT-01108 Vilnius, Lithuania }

\begin{abstract}
We propose a model of fractal point process driven by the
nonlinear stochastic differential equation. The model is adjusted
to the empirical data of trading activity in financial markets.
This reproduces the probability distribution function and power
spectral density of trading activity observed in the stock
markets. We present a simple stochastic relation between the
trading activity and return, which enables us to reproduce
long-range memory statistical properties of volatility by
numerical calculations based on the proposed fractal point
process.

\end{abstract}


\maketitle

\section{Introduction}

There are empirical evidences that the trading activity, the
trading volume and the volatility of the financial markets are
stochastic variables with the power-law probability distribution
function (PDF) \cite{Mandelbrot, Lux} and the long-range
correlations \cite{Engle, Plerou, Gabaix}. Most of proposed models
apply generic multiplicative noise responsible for the power-law
probability distribution function (PDF), whereas the long-range
memory aspect is not accounted in the widespread models
\cite{BookDacorogna}.The additive-multiplicative stochastic models
of the financial mean-reverting processes provide rich spectrum of
shapes for PDF, depending on the model parameters
\cite{Anteneodo}, however, do not describe the long-memory
features. Empirical analysis confirms that the long-range
correlations in volatility arise due to those of trading activity
\cite{Plerou}. On the other hand, trading activity is a financial
variable dependant on the one stochastic process, i.e.interevent
time between successive market trades. Therefore, it can be
modeled as event flow of the stochastic point process.

Recently, we investigated analytically and numerically the
properties of the stochastic multiplicative point processes,
derived formula for the power spectrum
\cite{GontisPhA,KaulakysPRE} and related the model with the
multiplicative stochastic differential equations
\cite{KaulakysPhA}. Preliminary comparison of the model with the
empirical data of the power spectrum and probability distribution
of stock market trading activity \cite{Gontis2} stimulated us to
work on the more detailed definition of the model. Here we present
the stochastic model of the trading activity with the long-range
correlations and investigate its connection to the stochastic
modeling of the volatility. The proposed stochastic nonlinear
differential equations reproduce the power spectrum and PDF of the
trading activity in the financial markets, describe the stochastic
interevent time as the fractal-based point process and can be
applicable for modeling of the volatility with the long-range
autocorrelation.

\section{Modeling fractal point process by the nonlinear stochastic differential equation}

Trades in financial markets occur at discrete times
${t_{1},t_{2},t_{1},...t_{k},...}$ and can be considered as
identical point events. Such point process is stochastic and
totaly defined by the stochastic interevent time
$\tau_{k}=t_{k+1}-t_{k}$. A fractal stochastic point process
results, when at least two statistics exhibit the power-law
scaling, indicating that represented phenomena contains clusters
of events over all scales of time \cite{Lowen}. The dimension of
the fractal point process is a measure of the clustering of the
events within the process and by the definition coincides with the
exponent of the power spectral density of the flow of events.

We can model the trading activity in financial markets by the
fractal point process as its empirical PDF and the power spectral
density exhibit the power-law scaling \cite{Plerou,Plerou2}. In
this paper we consider the flow of trades in financial markets as
Poisson process driven by the multiplicative stochastic equation.
First of all we define the stochastic rate $n=1/\tau$ of event
flow by continuous stochastic differential equation
\begin{equation}
\mathrm{d}\tau=\gamma\tau^{2\mu-2}\mathrm{d}t+\sigma\tau^{\mu-1/2}\mathrm{d}W,
\label{eq:taustoch}
\end{equation}
where $W$ is a standard random Wiener process, $\sigma$ denotes
the standard deviation of the white noise, $\gamma\ll 1$ is a
coefficient of the nonlinear damping and $\mu$ defines the power
of noise multiplicativity. The diffusion of $\tau$ should be
restricted at least from the side of high values. Therefore we
introduce an additional term
$-\frac{m}{2}\sigma^2\left(\frac{\tau}{\tau_{0}}\right)^m\tau^{2\mu-2}$
into the Eq. (\ref{eq:taustoch}), which produces the exponential
diffusion reversion in equation
\begin{equation}
\mathrm{d}\tau=\left[\gamma-\frac{m}{2}\sigma^2\left(\frac{\tau}{\tau_{0}}\right)^m\right]\tau^{2\mu-2}\mathrm{d}t+\sigma\tau^{\mu-1/2}\mathrm{d}W,
\label{eq:taustoch2}
\end{equation}
where $m$ and $\tau_{0}$ are the power and value of the diffusion
reversion, respectively. The associated Fokker-Plank equation with
the zero flow gives the simple stationary PDF
\begin{equation}
P(\tau)\sim\tau^{\alpha+1}\exp\left[-\left(\frac{\tau}{\tau_{0}}\right)^m\right]\label{eq:taudistrib}
\end{equation}
with  $\alpha=2(\gamma_{\sigma}-\mu)$ and
$\gamma_{\sigma}=\gamma/\sigma^2$. Eq. (\ref{eq:taustoch2})
describes continuous stochastic variable $\tau$, defines the rate
$n=1/\tau$ and, after the Ito transform of variable, results in
stochastic differential equation
\begin{equation}
\mathrm{d}n=\sigma^2\left[(1-\gamma_{\sigma})+\frac{m}{2}\left(\frac{n_{0}}{n}\right)^{m}\right]n^{2\eta-1}\mathrm{d}t+\sigma
n^{\eta}\mathrm{d}W, \label{eq:nstoch}
\end{equation}
where $\eta=\frac{5}{2}-\mu$ and $n_{0}=1/\tau_{0}$. Eq.
(\ref{eq:nstoch}) describes stochastic process $n$ with PDF
\begin{equation}
P(n)\sim\frac{1}{n^{\lambda}}\exp\left\{-\left(\frac{n_{\mathrm{0}}}{n}\right)^m
\right\},\quad\lambda=2(\eta-1+\gamma_{\sigma}),\label{eq:ndistr}
\end{equation}
and power spectrum $S(f)$ \cite{KaulakysPRE,KaulakysPhA,GontisPhA}
\begin{equation}
S(f)\sim\frac{1}{f^{\beta}},\quad\beta=2-\frac{3-2\gamma_{\sigma}}{2\eta-2}.
\label{eq:nspekt}
\end{equation}

Noteworthy, that in the proposed model  only two parameters,
$\gamma_{\sigma}$ and $\eta$ (or $\mu$), define exponents
$\lambda$ and $\beta$  of two power-law statistics, i.e., of PDF
and of the power spectrum. Time scaling parameter $\sigma^2$ in
Eq. (\ref{eq:nstoch}) can be omitted adjusting the time scale.
Here we define the fractal point process driven by the stochastic
differential equation (\ref{eq:nstoch}) or equivalently by Eq.
(\ref{eq:taustoch2}), i.e., we   assume $\tau(t)$ as slowly
diffusing mean interevent time of Poisson process with the
stochastic rate $n$. This should produce  the fractal point
process with the statistical properties defined by Eqs.
(\ref{eq:ndistr}) and (\ref{eq:nspekt}). Within this assumption
the conditional probability of interevent time $\tau_{\mathrm{p}}$
in the modulated Poisson point process with the stochastic rate
$1/\tau$ is
\begin{equation}
\varphi(\tau_{\mathrm{p}}|\tau)=\frac{1}{\tau}\exp\left[-\frac{\tau_{\mathrm{p}}}{\tau}\right].\label{eq:taupoisson}
\end{equation}
Then the long time distribution $\varphi(\tau_{\mathrm{p}})$ of
interevent time $\tau_{\mathrm{p}}$ has the integral form
\begin{equation}
\varphi(\tau_{\mathrm{p}})=C\int_{0}^{\infty}\exp\left[-\frac{\tau_{\mathrm{p}}}{\tau}\right]\tau^{\alpha}\exp\left[-\left(\frac{\tau}{\tau_{0}}\right)^m\right]\mathrm{d}
\tau,\label{eq:taupdistrib}
\end{equation}
with $C$ defined from the normalization,
$\int_{0}^{\infty}\varphi(\tau_{\mathrm{p}})\mathrm{d}
\tau_{\mathrm{p}}=1$. In the case of pure exponential diffusion
reversion, $m=1$, PDF (\ref{eq:taupdistrib}) has a simple form
\begin{equation}
\varphi(\tau_{\mathrm{p}})=\frac{2}{\Gamma(2+\alpha)\tau_{0}}\left(\frac{\tau_{\mathrm{p}}}{\tau_{0}}\right)
^{\frac{1+\alpha}{2}}\mathrm{K_{(1+\alpha)}}\left(2
\sqrt{\frac{\tau_{\mathrm{p}}}{\tau_{0}}}\right),\label{eq:taupintegr}
\end{equation}
where $\mathrm{K_{\alpha}}\left(z\right)$ denotes the modified
Bessel function of the second kind. For $m>1$ more complicated
structures of  distribution $\varphi(\tau_{\mathrm{p}})$ expressed
in terms of hypergeometric functions arise.

\section{Adjustment of the model to the empirical data}

We will investigate how the proposed modulated Poisson stochastic
point process  can be adjusted to the empirical trading activity,
defined as number of transactions in the selected time window
$\tau_{\mathrm{d}}$. Stochastic variable $n$ denotes the number of
events per unit time interval. One has to integrate the stochastic
signal Eq. (\ref{eq:nstoch}) in the time interval
$\tau_{\mathrm{d}}$ to get the number of events in the selected
time window. In this paper we denote the integrated number of
events as
\begin{equation}
N(t,\tau_{\mathrm{d}})=\int_{t}^{t+\tau_{\mathrm{d}}}n(t^{\prime})\mathrm{d}
t^{\prime}
\end{equation}
and call it the trading activity in the case of the financial
market.

 Detrended fluctuation analysis \cite{Plerou2} is one
of the methods to analyze the second order statistics related to
the autocorrelation of trading activity. The exponents $\nu$ of
the detrended fluctuation analysis  obtained by fits for each of
the 1000 US stocks show a relatively narrow spread of $\nu$ around
the mean value $\nu=0.85\pm0.01$ \cite{Plerou2}. We use relation
$\beta=2\nu-1$ between the exponents $\nu$ of detrended
fluctuation analysis and the exponents $\beta$ of the power
spectrum \cite{Beran} and in this way define the empirical value
of the exponent for the power spectral density $\beta=0.7$.

Our analysis of the Lithuanian stock exchange data confirmed that
the power spectrum of trading activity is the same for various
liquid stocks even for the emerging markets \cite{Gontis2}. The
histogram of exponents obtained by fits to the cumulative
distributions of trading activites of 1000 US stocks
\cite{Plerou2} gives the value of exponent $\lambda=4.4\pm0.05$
describing the power-law behavior of the trading activity.
Empirical values of $\beta=0.7$ and $\lambda=4.4$ confirm that the
time series of the trading activity in real markets are fractal
with the power law statistics. Time series generated by stochastic
process (\ref{eq:nstoch}) are fractal in the same sense.

Nevertheless, we face serious complications trying to adjust model
parameters to the empirical data of the financial markets. For the
pure multiplicative model, when $\mu=1$ or $\eta=3/2$, we have to
take $\gamma_{\sigma}=0.85$ to get $\beta=0.7$ and
$\gamma_{\sigma}=1.7$ to get $\lambda=4.4$, i.e. it is impossible
to reproduce the empirical PDF and power spectrum with the same
relaxation parameter $\gamma_{\sigma}$ and exponent of
multiplicativity $\mu$. We have proposed possible solution of this
problem in our previous publications \cite{GontisPhA,Gontis2}
deriving PDF for the trading activity $N$
\begin{equation}
P(N)\sim\left\{
\begin{array}{ll}
\frac{1}{N^{3+\alpha}},& N\ll \gamma^{-1}, \\
\frac{1}{N^{5+2\alpha}},& N\gg \gamma^{-1}.
\end{array}
\right. \label{eq:Ndistrib}
\end{equation}

When $N\gg \gamma^{-1}$ this yields exactly the required value of
$\lambda=5+2\alpha=4.4$ and $\beta=0.7$ for
$\gamma_{\sigma}=0.85$.

Nevertheless, we cannot accept this as the sufficiently accurate
model of the trading activity since the empirical power law
distribution is achieved only for very high values of the trading
activity. Probably this reveals the mechanism how the power law
distribution converges to normal distribution through the growing
values of the exponent, but empirically observed power law
distribution in wide area of $N$ values cannot be reproduced. Let
us notice here that the desirable power law distribution of the
trading activity with the exponent $\lambda=4.4$ may be generated
by the model (\ref{eq:nstoch}) with  $\eta=5/2$ and
$\gamma_{\sigma}=0.7$. Moreover, only the smallest values of
$\tau$ or high values of $n$ contribute to the power spectral
density of trading activity \cite{KaulakysPhA}. This suggests us
to combine the stochastic process with two values of $\mu$: (i)
$\mu\simeq0$ for the main area of $\tau$ and $n$ diffusion  and
(ii) $\mu=1$ for the lowest values of $\tau$ or highest values of
$n$. Therefore, we introduce a new stochastic differential
equation for $n$ combining two powers of the multiplicative noise,

\begin{equation}
\mathrm{d}
n=\sigma^2\left[(1-\gamma_{\sigma})+\frac{m}{2}\left(\frac{n_{0}}{n}\right)^{m}\right]\frac{n^4}{(n\epsilon+1)^2}\mathrm{d}
t+\frac{\sigma n^{5/2}}{(n\epsilon+1)}\mathrm{d}W,
\label{eq:nstoch2}
\end{equation}
where a new parameter $\epsilon$ defines crossover between two
areas of $n$ diffusion. The corresponding iterative equation for
$\tau_{k}$ in such a case is

\begin{equation}
\tau_{k+1}=\tau_{k}+\left[\gamma-\frac{m}{2}\sigma^2\left(\frac{\tau}{\tau_{0}}\right)^m\right]\frac{\tau_{k}}{(\epsilon+\tau_{k})^2}+\sigma\frac{\tau_{k}}{\epsilon+\tau_{k}}\varepsilon_{k},
\label{eq:tauiterat2}
\end{equation}
where $\varepsilon_{k}$ denotes uncorrelated normally distributed
random variable with the zero expectation and unit variance.

\begin{figure}
\begin{center}
\includegraphics[width=.4\textwidth]{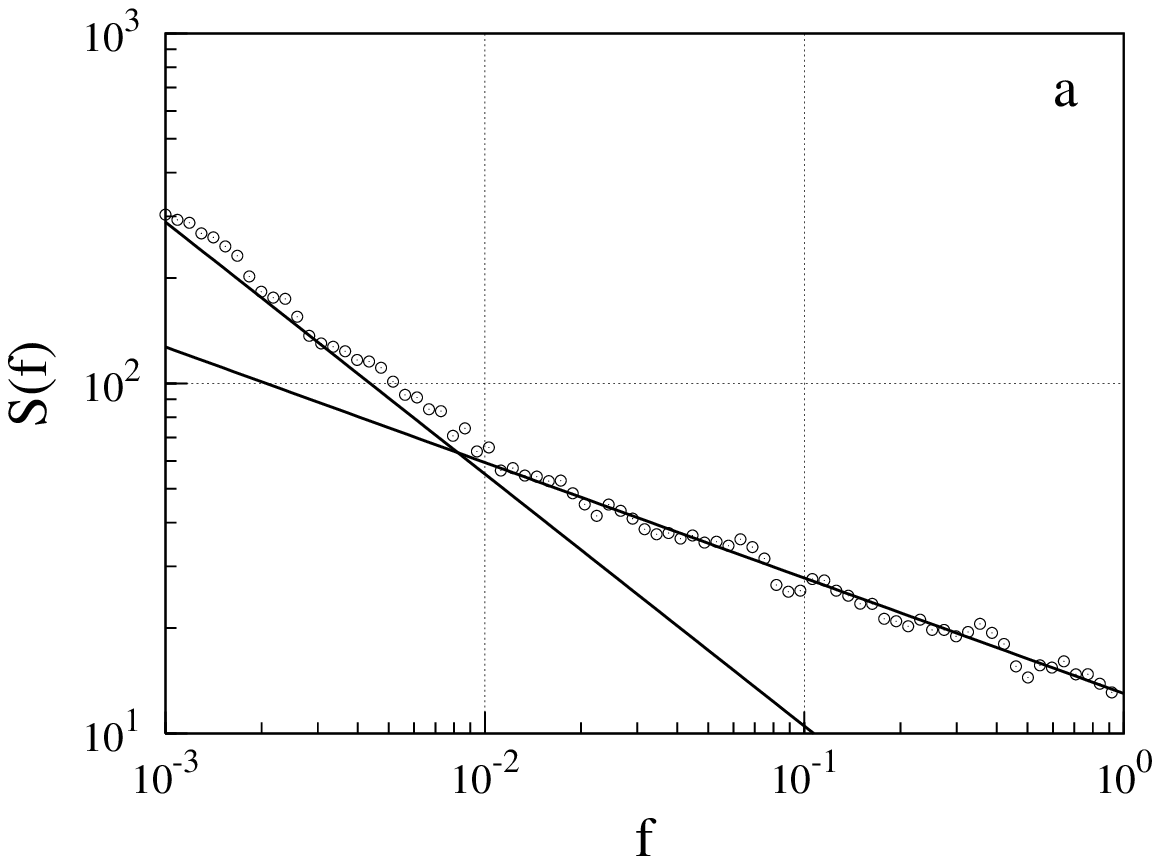}
\includegraphics[width=.4\textwidth]{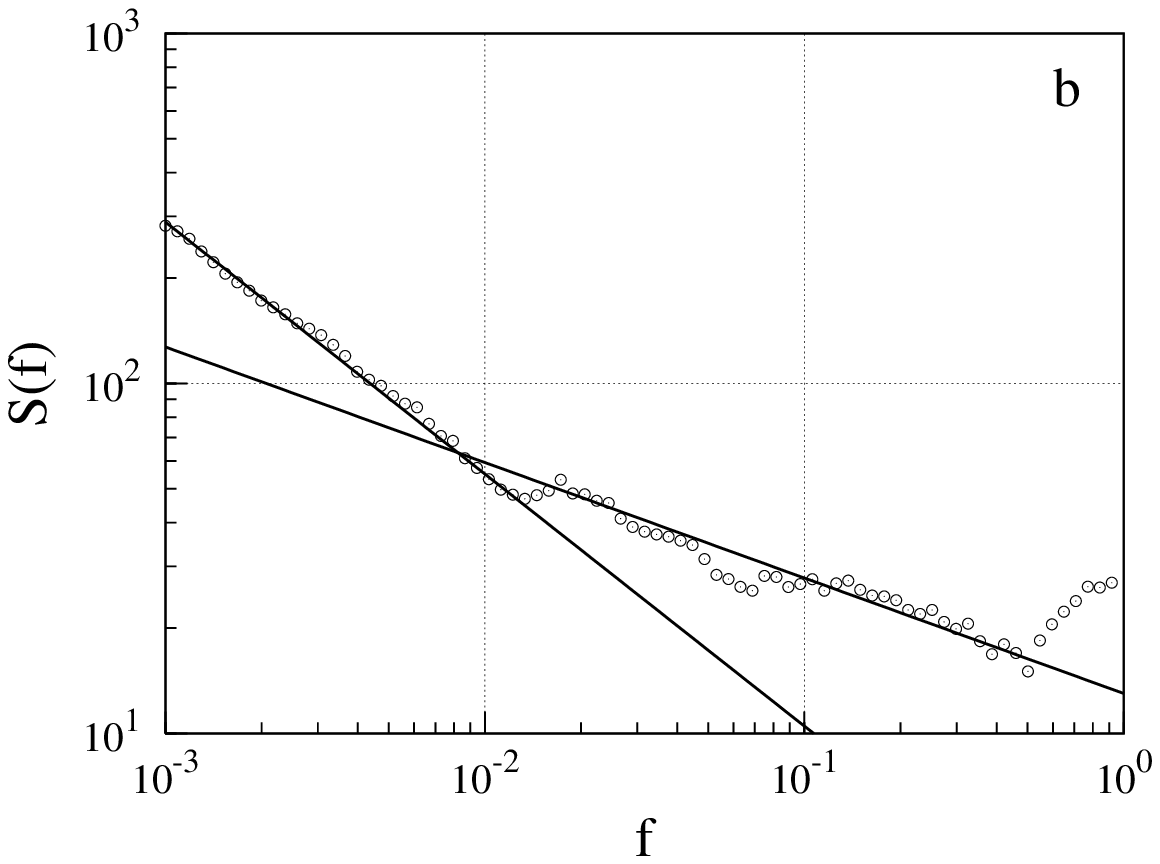}
\end{center}
\caption{Power spectral density $S(f)$ averaged over 100
realisations of series with 1000000 iterations and parameters
$\gamma=0.0004$; $\sigma=0.025$; $\epsilon=0.07$; $\tau_{0}=1$;
$m=6$. Straight lines approximate power spectrum
$S\sim1/f^{\beta_{1,2}}$ with $\beta_{1}=0.33$ and
$\beta_{2}=0.72$: a) $S(f)$ of the flow
$I(t)=\sum_{k}\delta(t-t_{k})$ with the interevent time
$\tau_{k}=t_{k+1}-t_{k}$ generated by Eq. (\ref{eq:tauiterat2}),
b) $S(f)$ calculated by the Fast Fourier Transform of $n$ series
generated by Eq. (\ref{eq:nstoch2}).} \label{fig:1}
\end{figure}

Eqs. (\ref{eq:nstoch2}) and (\ref{eq:tauiterat2}) define related
stochastic variables $n$ and $\tau$, respectively, and they should
reproduce the long-range statistical properties of the trading
activity and of waiting time in the financial markets. We verify
this by the numerical calculations. In figure~\ref{fig:1} we
present the power spectral density calculated for the equivalent
processes (\ref{eq:nstoch2}) and (\ref{eq:tauiterat2}) (see
\cite{GontisPhA} for details of calculations). This approach
reveals the structure of the power spectral density in wide range
of frequencies and shows that the model exhibits not one but
rather two separate power laws with the exponents $\beta_{1}=0.33$
and $\beta_{2}=0.72$. From many numerical calculations performed
with the multiplicative point processes we can conclude that
combination of two power laws of spectral density arise only when
the multiplicative noise is a crossover of two power laws as in
Eqs. (\ref{eq:nstoch2}) and (\ref{eq:tauiterat2}). We will show in
the next section that this may serve as an explanation of two
exponents of the power spectrum in the empirical data of
volatility for \verb"S&P 500" companies \cite{Liu}.

\begin{figure}
\begin{center}
\includegraphics[width=.8\textwidth]{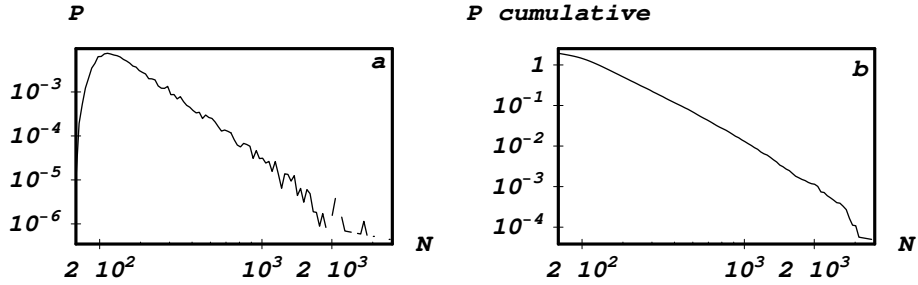}
\hspace{-10pt}
\end{center}
\par
\vspace{-10pt} \caption{a) PDF $P(N)$ calculated from the
histogram of $N$ generated by Eq. (\ref{eq:tauiterat2}) with the
selected time interval $\tau_{\mathrm{d}}=100$. b) cumulative
distribution $P_{>}(N)$. Other parameters are as in
figure~\ref{fig:1}.} \label{fig:2}
\end{figure}

Empirical data of the trading activity statistics should be
modeled by the integrated flow of events $N$ defined in the time
interval $\tau_{\mathrm{d}}\gg\tau_{0}$. In figure~\ref{fig:2} we
demonstrate the probability distribution functions $P(N)$ and its
cumulative form $P_{>}(n)$ calculated from the histogram of $N$
generated by Eq. (\ref{eq:tauiterat2}) with the selected time
interval $\tau_{\mathrm{d}}=100$. This illustrates that the model
distribution of the integrated signal $N$ has the power-law form
with the same exponent $\lambda=4.4$ as observed in empirical data
\cite{Plerou,Gabaix}.

The power spectrum of the trading activity $N$ has the same
exponent $\beta=0.7$ as power spectrum of $n$ in the low frequency
area for all values of $\tau_{\mathrm{d}}$.

The same numerical results can be reproduced by continuous
stochastic differential equation (\ref{eq:nstoch2}) or iteration
equation (\ref{eq:tauiterat2}). One can consider the discrete
iterative equation for the interevent time $\tau_{k}$
(\ref{eq:tauiterat2}) as a method to solve numerically continuous
equation

\begin{equation}
\mathrm{d}\tau=\left[\gamma-\frac{m}{2}\sigma^2\left(\frac{\tau}{\tau_{0}}\right)^m\right]\frac{1}{(\epsilon+\tau)^2}\mathrm{d}t
+\sigma\frac{\sqrt{\tau}}{\epsilon+\tau}\mathrm{d}W.
\label{eq:taucontinuous}
\end{equation}
The continuous equation (\ref{eq:nstoch2}) follows from the Eq.
(\ref{eq:taucontinuous}) after change of variables $n=1/\tau$.

\begin{figure}
\begin{center}
\includegraphics[width=.8\textwidth]{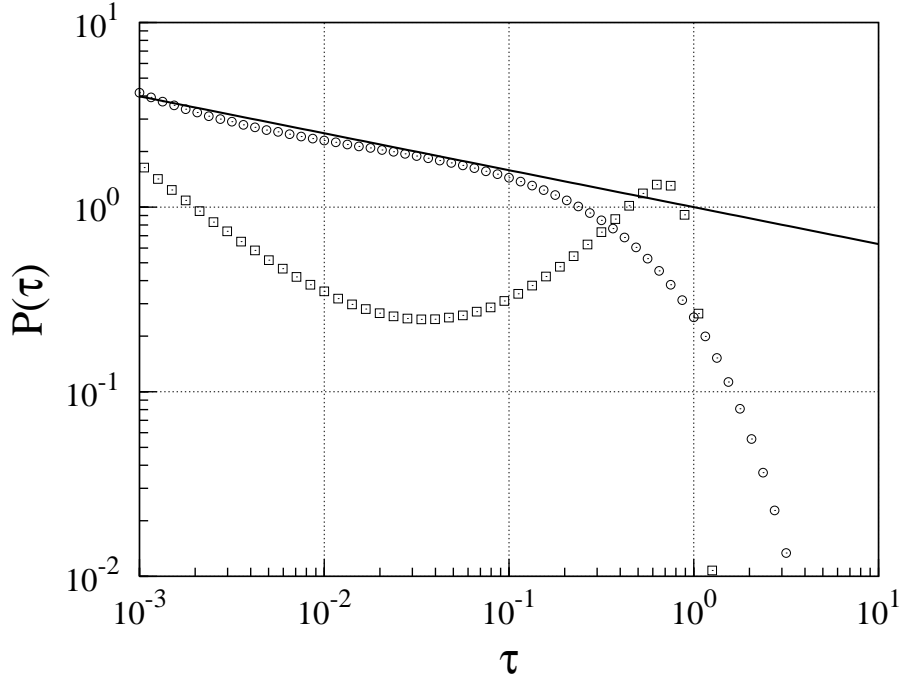}
\hspace{-10pt}
\end{center}
\par
\vspace{-10pt} \caption{PDF of interevent time $P(\tau)$: open
circles, calculated from the histogram of $\tau_{\mathrm{p}}$
generated by Eq. (\ref{eq:taupoisson}) with the mean interevent
time calculated from Eq. (\ref{eq:taucontinuous}); open squares,
calculated form the iterative equation (\ref{eq:tauiterat2}). Used
parameters are as in figure 1. Straight line approximates power
law $P(\tau_{\mathrm{p}})\sim\tau_{\mathrm{p}}^{-0.2}$. }
\label{fig:3}
\end{figure}

We can conclude that the long-range memory properties of the
trading activity in the financial markets as well as the PDF can
be modeled by the continuous stochastic differential equation
(\ref{eq:nstoch2}). In this model the exponents of the power
spectral density, $\beta$, and of PDF, $\lambda$, are defined by
one parameter $\gamma_{\sigma}=\gamma/\sigma^{2}$. We consider the
continuous equation of the mean interevent time $\tau$ as a model
of slowly varying stochastic rate $1/\tau$ in the modulated
Poisson process (\ref{eq:taupoisson}). In figure~\ref{fig:3} we
demonstrate the probability distribution functions
$P(\tau_{\mathrm{p}})$ calculated from the histogram of
$\tau_{\mathrm{p}}$ generated by Eq. (\ref{eq:taupoisson}) with
the diffusing mean interevent time calculated from Eq.
(\ref{eq:taucontinuous}).

Numerical results show good qualitative agreement with the
empirical data of interevent time probability distribution
measured from few years series of U.S. stock data \cite{Ivanov}.
This enables us to conclude that the proposed stochastic model
captures the main statistical properties including PDF and the
long-range correlation of the trading activity in the financial
markets. Furthermore, in the next section we will show  that this
may serve as a background statistical model responsible for the
statistics of return volatility in widely accepted geometric
Brownian motion (GBM) of the financial asset prices.

\section{Modeling long-range memory volatility}

The basic quantities studied for the individual stocks are price
$p(t)$ and return

\begin{equation}
x(t,\tau_{\mathrm{d}})=\ln p(t+\tau_{\mathrm{d}})-\ln p(t)
\label{eq:return}
\end{equation}

Let us express return $x(t,\tau_{\mathrm{d}})$ over a time
interval $\tau_{\mathrm{d}}$ through the subsequent changes
$\delta x_{i}$ due to the trades $i=1,2....N(t,\tau_{\mathrm{d}})$
in the time interval $[t,t+\tau_{\mathrm{d}}]$,

\begin{equation}
x(t,\tau_{\mathrm{d}})=\sum_{i=1}^{N(t,\tau_{\mathrm{d}})}\delta
x_{i}. \label{eq:return2}
\end{equation}

We denote the variance of $\delta x_{i}$ calculated over the time
interval $\tau_{\mathrm{d}}$ as $W^{2}(t,\tau_{\mathrm{d}})$. If
$\delta x_{i}$ are mutually independent one can apply the central
limit theorem to sum (\ref{eq:return2}). This implies that for the
fixed variance $W^{2}(t,\tau_{\mathrm{d}})$ return
$x(t,\tau_{\mathrm{d}})$ is a normally distributed random variable
with the variance
$W^{2}(t,\tau_{\mathrm{d}})N(t,\tau_{\mathrm{d}})$

\begin{equation}
x(t,\tau_{\mathrm{d}})=W(t,\tau_{\mathrm{d}})\sqrt{N(t,\tau_{\mathrm{d}})}\varepsilon_{t}
, \label{eq:return3}
\end{equation}
where $\varepsilon_{t}$ is the normally distributed random
variable with the zero expectation and unit variance.

\begin{figure}
\begin{center}
\includegraphics[width=.6\textwidth]{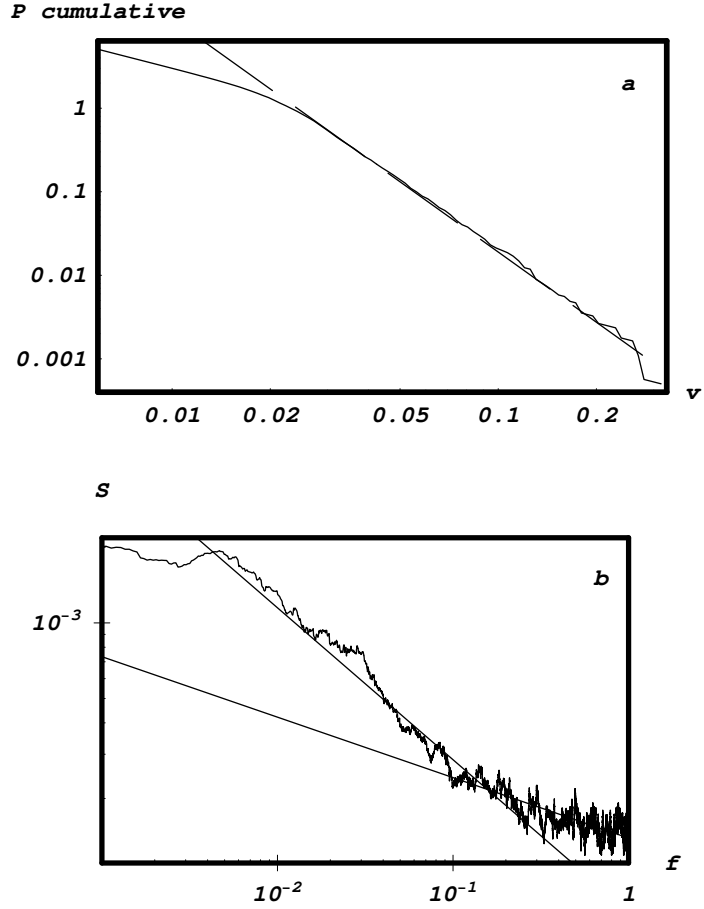}
\hspace{-10pt}
\end{center}
\par
\vspace{-10pt} \caption{ (a) Cumulative probability distribution
function of the volatility, $P_{>}(\overline{v})$, averaged over
10 intervals calculated from the series of $n(t)$ generated by
Eqs. (\ref{eq:nstoch2}) and (\ref{eq:return4}), all parameters are
the same as in previous calculations.  Dashed line approximates
the power law $P(\overline{v})\sim 1/\overline{v}^{2.8}$. (b)
Power spectral density $S(f)$ of $v$ calculated from FFT of the
same series $n(t)$. Straight lines approximate power spectral
density $S\sim 1/f^{\beta_{1,2}}$ with $\beta_{1}=0.6$ and
$\beta_{2}=0.24$.} \label{fig:4}
\end{figure}

Empirical test of conditional probability
$P(x(t,\tau_{\mathrm{d}})|W(t,\tau_{\mathrm{d}}))$ \cite{Plerou}
confirms its Gaussian form, and the unconditional distribution
$P(x(t,\tau_{\mathrm{d}}))$ is a power-law with the cumulative
exponent $3$. This implies that the power-law tails of returns are
largely due to those of $W(t,\tau_{\mathrm{d}})$. Here we refer to
the theory of price diffusion as a mechanistic random process
\cite{Farmer,Farmer2}. For this idealized model the short term
price diffusion depends on the limit order removal and this way is
related to the market order flow. Furthermore, the empirical
analysis confirms that the volatility calculated for the fixed
number of transactions has the long memory properties as well and
it is correlated with real time volatility \cite{Farmer3}. We
accumulate all these results into the assumption that standard
deviation $W(t,\tau_{\mathrm{d}})$ may be proportional to the
square root of the trading activity, i.e.,
$W(t,\tau_{\mathrm{d}})\sim k \sqrt{N(t,\tau_{\mathrm{d}})}$. This
enables us to propose a simple model of return

\begin{equation}
x(t,\tau_{\mathrm{d}})\sim k N(t,\tau_{\mathrm{d}})\varepsilon_{t}
\label{eq:return4}
\end{equation}
and related model of volatility $v=|x(t,\tau_{\mathrm{d}})|$ based
on the proposed model of trading activity (\ref{eq:nstoch2}). We
generate series of trade flow $n(t)$ numerically solving Eq.
(\ref{eq:nstoch2}) with variable steps of time $\Delta
t_{i}=h_{i}=n_{0}/n_{i}$ and calculate the trading activity in
subsequent time intervals $\tau_{\mathrm{d}}$ as
$N(t,\tau_{\mathrm{d}})=\int_{t}^{t+\tau_{\mathrm{d}}}n(t^{\prime})dt^{\prime}$.
This enables us to generate series of return
$x(t,\tau_{\mathrm{d}})$, of volatility
$v(t,\tau_{\mathrm{d}})=|x(t,\tau_{\mathrm{d}})|$ and of the
averaged volatility
$\overline{v}=\frac{1}{m}\sum_{i=1}^{i=m}v(t_{i},\tau_{\mathrm{d}})$.

In figure~\ref{fig:4} we demonstrate cumulative distribution of
$\overline{v}$ and the power spectral density of
$v(t,\tau_{\mathrm{d}})$ calculated from FFT. We see that proposed
model enables us to catch up the main features of the volatility:
the power law distribution with exponent $2.8$ and power spectral
density with two exponents $\beta_{1}=0.6$ and $\beta_{2}=0.24$.
This is in a good agreement with the empirical data
\cite{Liu,Farmer3}.







\section{\label{sec:concl}Conclusions}

Starting from the concept of the fractal point processes
\cite{Lowen} we proposed process driven by the nonlinear
stochastic differential equation and based on the earlier
introduced stochastic point process model
\cite{KaulakysPRE,GontisPhA,KaulakysPhA,Gontis2}. This may serve
as a possible model of the flow of points or events in the
physical, biological and social systems when their statistics
exhibit power-law scaling indicating that the represented
phenomena contains clusters of events over all scales. First of
all, we analyze the statistical properties of trading activity and
waiting time in financial markets by the proposed Poisson process
with the stochastic rate defined as a stand-alone stochastic
variable. We consider the stochastic rate as continuous one and
model it by the stochastic differential equation, exhibiting
long-range memory properties \cite{KaulakysPhA}. Further we
propose a new form of the stochastic differential equation
combining two powers of multiplicative noise: one responsible for
the probability distribution function and another responsible for
the power spectral density. The proposed new form of the
continuous stochastic differential equation enabled us to
reproduce the main statistical properties of the trading activity
and waiting time, observable in the financial markets. In the new
model the power spectral density with two different scaling
exponents arise. This is in agreement with the empirical power
spectrum of volatility and implies that the market behavior may be
dependant on the level of activity. One can observe at least two
stages in market behavior: calm and excited. Finally,  we propose
a very simple stochastic relation between trading activity and
return to reproduce the statistical properties of volatility. This
enabled us to model empirical distribution and long-range memory
of volatility.

\noindent \textbf{Acknowledgment}

This work was supported by the Lithuanian State and Studies
Foundation. The authors thank Dr. M. Alaburda for kind assistance
preparing illustrations.


\begin{thebibliography}{9}
\setlength{\itemsep}{-2pt}

\bibitem{Mandelbrot} Mandelbrot B B, J.Business {\bf 36}, 394 (1963).

\bibitem{Lux} Lux T, Appl.Fin. Econ. {\bf 6}, 463 (1996).

\bibitem{Engle} Engle R. F.  and Patton A. J., Quant. Finance {\bf 1}, 237 (2001).

\bibitem{Plerou} Plerou V, Gopikrishnan P, Gabaix X, Amaral L A N and Stanley H E, 2001 Quant.
Finance {\bf 1} 262

\bibitem{Gabaix} Gabaix X, Gopikrishnan P, Plerou V,  Stanley H E,
2003 Nature {\bf 423} 267

\bibitem{BookDacorogna} Dacorogna M M, Gencay R, Muller U A,
Olsen R B and Pictet O V, 2001 {\em An Introduction to
High-Frequency Finance} (Academic Press, San Diego)

\bibitem{Anteneodo} Anteneodo C and Riera R, 2005 Phys. Rev. E {\bf 72}
026106

\bibitem{KaulakysPRE} Kaulakys B, Gontis V and Alaburda M, 2005 Phys. Rev. E
{\bf71} 051105

\bibitem{GontisPhA} Gontis V and Kaulakys B, 2004 Physica A {\bf
343} 505

\bibitem{KaulakysPhA} Kaulakys B, Ruseckas J, Gontis V and Alaburda
M, 2006 Physica A {\bf 365} 217

\bibitem{Gontis2} Gontis V and Kaulakys B, 2004 Physica A {\bf344}
128

\bibitem{Lowen} Lowen S B, Teich M C, 2005 {\em Fractal-Based Point Processes} (Wiley, ISBN: 0-471-38376-7)

\bibitem{Plerou2} Plerou V  E, Gopikrishnan P, Amaral L, Gabaix X and Stanley E, 2000 Phy. Rev. E {\bf 62} R3023

\bibitem{Beran} Beran J, 1994 {\em Statistics for Long-Memory Processes} (Chapman and Hall,
NY)

\bibitem{Liu} Liu Y, Gopikrishnan P, Cizeau P, Meyer M, Peng Ch
and Stanley H E, 1999 Phys. Rev. E {\bf60} 1390

\bibitem{Ivanov} Ivanov P, Yuen A, Podobnik B, Lee Y, 2004 Phys. Rev. E {\bf69}
056107

\bibitem{Farmer} Daniels M, Farmer D, Gillemot L, Iori G and
Smith E, 2003 Phys. Rev. Lett. {\bf90} 108102

\bibitem{Farmer2} Farmer D, Gillemot L, Lillo F, Szabolcs M and Sen A, 2004 Quantative Finance {\bf4}
383

\bibitem{Farmer3} Gillemot L, Farmer J D, Lillo F, Santa Fe Institute Working Paper 05-12-041

\end{thebibliography}
\end{document}